\documentclass[12pt]{iopart}

\usepackage[T1]{fontenc}
\usepackage{color}
 \expandafter\let\csname equation*\endcsname\relax
 \expandafter\let\csname endequation*\endcsname\relax
\usepackage{amsmath}
\usepackage{bbm}
\input epsf
\usepackage{iopams}  

\usepackage[pdftex]{graphicx}

\begin{document}

\title[]{Superconductivity in the three-band model of cuprates: nodal direction characteristics and influence of intersite interactions}

\author{M Zegrodnik$^1*$, A Biborski$^1$, M. Fidrysiak$^{2}$, and J Spa\l ek$^{2}$}

\address{$^1$Academic Centre for Materials and Nanotechnology, AGH University of Science and Technology, Al. Mickiewicza 30, 30-059 Krak\'ow\\
$^2$Institute of Theoretical Physics, Jagiellonian University, {\L}ojasiewicza  11, 30-348 Krak\'ow, Poland}

\ead{michal.zegrodnik@agh.edu.pl}

\vspace{10pt}
\begin{indented}
\item[]September 2020
\end{indented}

\begin{abstract}
The three-band Emery model is applied to study the selected principal features of the $d$-$wave$ superconducting phase in the copper-based compounds. The electron-electron correlations are taken into account by the use of the diagrammatic expansion of the Guztwiller wave function (DE-GWF method). The nodal Fermi velocity, Fermi momentum, and effective mass are all determined in the paired state and show relatively good agreement with the available experimental data, as well as with the corresponding single-band calculations. Additionally, the influence of the next-nearest neighbor oxygen-oxygen hopping and intersite Coulomb repulsion terms on the superconducting phase is analyzed.
\end{abstract}

%
%
%
%
%

The superconducting state in the copper-based compounds has long been the subject of intense study \cite{cuprates_rev_2006,tJmodel_rev_2008}. Due to significant electron-electron repulsion appearing within the copper-oxygen planes of the cuprates, those materials belong to the group of the so-called strongly correlated electron systems. Calculation methods dedicated for such compounds are involved and their application is limited only to relatively simplified models. A significant effort has been devoted to the theoretical description of the copper-oxygen planes which are common to the whole cuprate family and are believed to be instrumental for the formation of the paired phase \cite{Anderson1988,cuprates_rev_2006,ZRS1988}. In this respect, the simplest approach is based on a single-band picture with the copper and oxygen degrees of freedom combined within the so-called Zhang-Rice singlets, which play the role of quasiparticles \cite{ZRS1988}. Along these lines, both the Hubbard and $t$-$J$ models have been investigated \cite{tJmodel_rev_2008,Dagotto1994} by means of various calculation techniques, taking into account the electron-electron correlation effects on different accuracy levels \cite{VMC_Hubbard_2007,DMFT_SC_Hub_2006,Hub_SC_2005,Gros2007,Fidrysiak2020}. The $t$-$J$-$U$ model, which combines the features of both Hubbard and $t$-$J$ models, has been extensively investigated by us recently \cite{Zegrodnik_1,Zegrodnik_5,Fidrysiak2018} and leads to very good semi-quantitative agreement between our theoretical results and principal experimental data \cite{Zegrodnik_1}.

In order to make the model more realistic and, at the same, time be able to take into account the electron-electron correlations, the three band Emery model (or $d$-$p$ model) has also been studied \cite{3band_PRL_Mott,QMC_3band_2016,Yanagisawa2008,Weber_epl2012,Zegrodnik_3band}. Within such approach, the $2p_x$ and $2p_y$ orbitals due to oxygen are explicitly taken into account in addition to $3d_{x^2-y^2}$ copper states. The experimental indications of the significant role of the oxygen degrees of freedom played in the physics of hole-doped cuprates has been provided in Refs. \cite{Zheng1995,Rybicki2016,Weber_epl2012,Ruan_Sci_Biull2016}. Recently, we have applied the variational approach based on the correlated Gutzwiller- and Jastrow-type wave functions to analyze both the selected normal-state characteristics and the paired state within the three-band ($d$-$p$) model, as well as to compare them with the corresponding single-band calculations \cite{Zegrodnik_3band,biborski2020,Zegrodnik_nematic}. The results of that analysis support the view that in general aspects the single- and three- band descriptions of electron-correlation-induced superconductivity lead to similar overall physical picture. Nevertheless, an explicit inclusion of the oxygen degrees of freedom seems to be necessary in order to carry out a detailed description of individual compounds from the cuprate family and in particular, to reconstruct the significantly different values of the maximal critical temperature in different systems \cite{Zegrodnik_3band}.

Here, we extend the analysis provided in Refs. \cite{Zegrodnik_3band} and compare the nodal direction characteristics, extracted from the three-band model, with the available experimental data, as well discuss their relation to the single-band calculations carried out earlier. We focus on the doping dependencies of the nodal Fermi velocity, nodal Fermi momentum and effective mass, which are among the principal features characterising the cuprates. It should be noted, that different variants of the $d$-$p$ model have been considered over the years; in most cases the inter-site Coulomb repulsion terms, as well as next-nearest neighbor hoppings were both neglected. However, recently it has been noted that the latter lead not only to quantitative renormalization of quasiparticle dynamics, but may be essential for describing qualitatively appical hybridization effects on magnetic fluctuations \cite{Peng2017}. Sensitivity of the maximal superconducting transition temperature on the long range hopping has been also pointed out previously in model studies \cite{Pavarini}. Finally, the long-range Coulomb interaction is believed to be necessary to correctly reproduce charge dynamics of cuprate superconductors which, in turn, may affect local pair formation \cite{Hepting,Greco}. Therefore, it is vital to analyze, apart from the nodal direction characteristics, also the influence of the mentioned additional terms, on the paired state. Explicitly, we consider the model with next-nearest neighbor oxygen-oxygen hopping and the intersite $d$-$d$, $d$-$p$ and $p_x$-$p_y$ Coulomb repulsion terms. Our aim is to evaluate the importance of the additional interactions, particularly those, which do not appear in the effective single-band picture. The additional motivation in the second part of our analysis is to verify if the overall features of the SC state are reconstructed within the extended $d$-$p$ model for the parameters obtained within the recently reported $ab$-$initio$ procedure which does not suffer from the double counting problem \cite{Hirayama_GWDFT_2018}.

In the following Section we present the details of the theoretical model and the applied computational methods. In Section III we analyze the nodal direction characteristics in the paired state and compere them with the available experimental data. Next, we study the influence of the next-nearest neighbor oxygen-oxygen hopping as well as the intersite Coulomb repulsion terms on the superconducting state. The conclusions are deferred to Section IV.

\section{Model and method}
\begin{figure}
 \centering
 \includegraphics[width=0.4\textwidth]{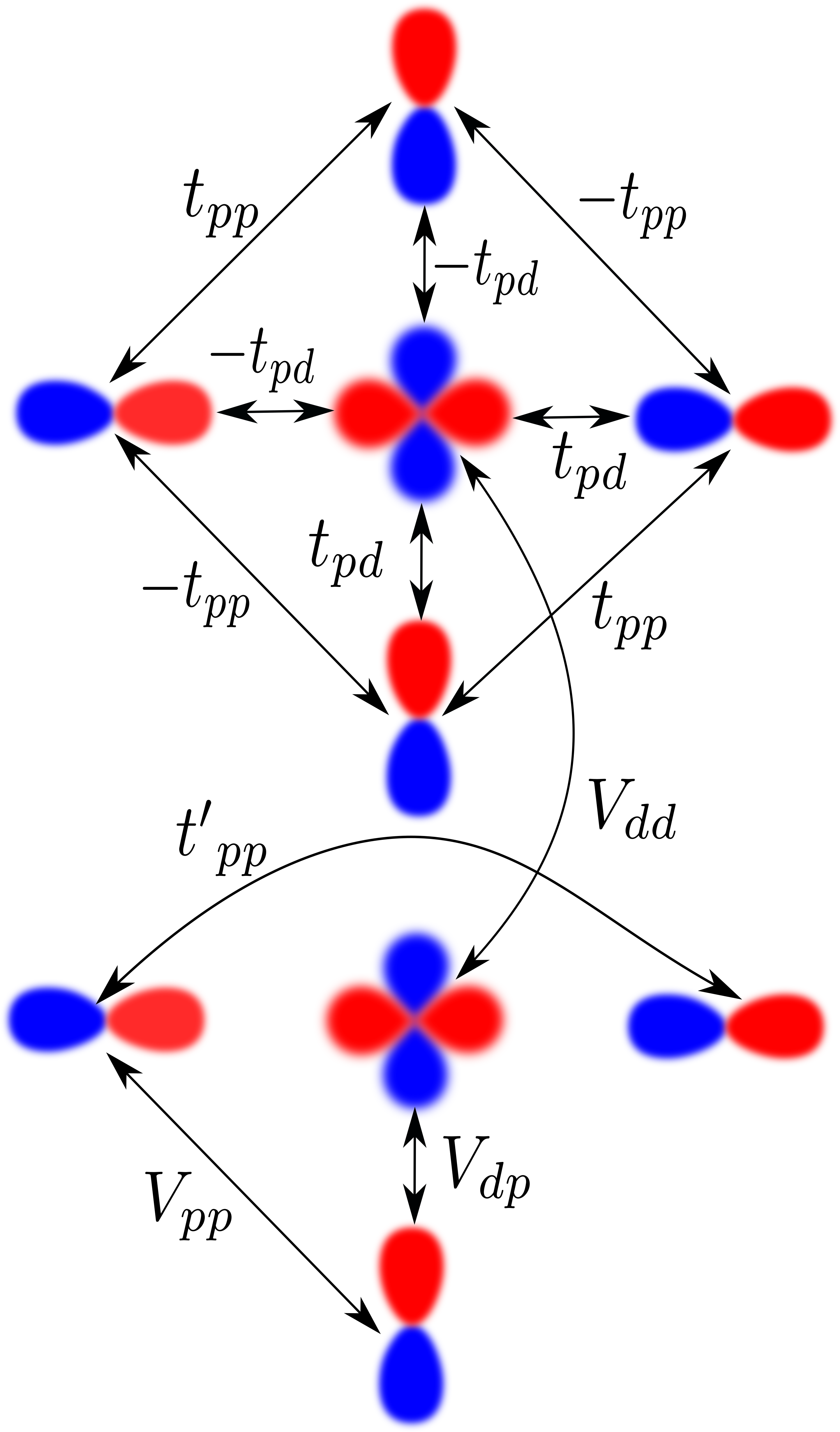}
 \caption{The hopping parameters between the three types of orbitals in the model and the corresponding sign convention for the antibonding orbital structure. The $d_{x^2-y^2}$  orbital is centered at the copper site and the $p_x/p_y$ orbitals are centered at the oxygen sites. Additionally, in the bottom part we show the next nearest-neighbors oxygen-oxygen hopping and the inter-site Coulomb repulsion terms which are taken into account in the second part of our analysis.}
 \label{fig:Cu_O_hoppings}
\end{figure}

The three-band $d$-$p$ model has the form
\begin{equation}
\begin{split}
 \hat{H}&=\sideset{}{'}\sum_{il,jl'}t^{ll'}_{il}\hat{c}^{\dagger}_{il\sigma}\hat{c}_{jl'\sigma}+\sum_{il}(\epsilon_{l}-\mu)\hat{n}_{il}+\sum_{il}U_{l}\hat{n}_{il\uparrow}\hat{n}_{il\downarrow}\\ 
 &+\sideset{}{''}\sum_{ijll'\sigma\sigma'}V_{ll'}\hat{n}_{il\sigma}\hat{n}_{jl'\sigma'},
 \end{split}
 \label{eq:Hamiltonian_start}
\end{equation}
where $\hat{c}^{\dagger}_{il\sigma}$ and $\hat{c}_{il\sigma}$ creates and anihilates electron with spin $\sigma$ at the $i$-th atomic site and orbital denoted by $l\in\{{d,p_x,p_y}\}$. The primed summation is carried out over the interorbital nearest neighbors as well as the intraorbital $p_x$-$p_x$ and $p_y$-$p_y$ next-nearest neighbors (cf. Fig. \ref{fig:Cu_O_hoppings}). The phase convention of the wave function determining the signs of the respective hopping parameters has been taken in the electron representation as shown in Fig. \ref{fig:Cu_O_hoppings}. The second term of the Hamiltonian introduces the $d$ and $p_x/p_y$ atomic levels ($\epsilon_{p_x}=\epsilon_{p_y}\equiv\epsilon_{p}$) with the energy shift $\epsilon_d-\epsilon_p\equiv\epsilon_{dp}$ between them. The chemical potential is denoted by $\mu$ and the interaction parameters $U_d$ and $U_{p_x}=U_{p_y}\equiv U_p$ correspond to the intrasite Coulomb repulsion between two electrons with opposite spins located on the $d$ and $p_x/p_y$ orbitals, respectively. Additionally, in the last term we take into account the intersite nearest neighbor $d$-$p$, $p_x$-$p_y$, and $d$-$d$ Coulomb repulsion terms denoted by $V_{dp}$, $V_{pp}$, and $V_{dd}$, respectively.

Due to the significant strength of the Coulomb repulsion at the $d$-orbitals, the electron-electron correlations are expected to significantly influence the physical features of the system. In order to take the correlation effects into account we use the approach based on the diagrammatic expansion of the Gutzwiller wave function (DE-GWF), which has been discussed by us extensively and applied both to single- and multi-band models \cite{Zegrodnik_1,Kaczmarczyk_2013,MWysokinski_Anderson,2band_Hub_Bunemann,Zegrodnik_3band}. Within the DE-GWF approach we use the Gutzwiller-like projected many particle wave function of the form
\begin{equation}
 |\Psi_G\rangle\equiv\hat{P}|\Psi_0\rangle=\prod_{il}\hat{P}_{il}|\Psi_0\rangle \;,
 \label{eq:GWF}
\end{equation}
where $|\Psi_0\rangle$ corresponds to uncorrelated state of the system and the projection operator is the following
\begin{equation}
 \hat{P}_{il}\equiv \sum_{\Gamma}\lambda_{\Gamma|il}|\Gamma\rangle_{il\;il}\langle\Gamma|\;,
 \label{eq:P_Gamma}
\end{equation}
where $\lambda_{\Gamma|il}$ are the variational parameters determining relative weights corresponding to $|\Gamma\rangle_{il}$, which in turn represent states of the local basis on the atomic sites with the three types of orbitals ($l\in\{{d,p_x,p_y}\}$)
\begin{equation}
|\Gamma\rangle_{il}\in \{|\varnothing\rangle_{il}, |\uparrow\rangle_{il}, |\downarrow\rangle_{il},
|\uparrow\downarrow\rangle_{il}\}\;.
\label{eq:local_states}
\end{equation}
The consecutive states represent the empty, singly, and doubly occupied local configurations, respectively. 
The energy minimization over the variational parameters allows to suppress the weight of electronic configurations that lead to increased interaction energies. More details regarding the application of the DE-GWF approach to the three band model of cuprates is provided in Ref. \cite{Zegrodnik_3band}. As shown there, due to low value of $U_p$ with respect to $U_d$, it is justified to omit the projection procedure at the oxygen sites and take $\lambda_{\Gamma|ip_x}=\lambda_{\Gamma|ip_y}\equiv 1$. Such simplification does not lead to significant changes in the calculated characteristics of the system as shown in Ref. \cite{Zegrodnik_3band}.

Within our analysis, the paired state is recognized by nonzero values of the so-called correlated pairing amplitudes (correlated gaps). As shown in Ref. \cite{Zegrodnik_3band}, a number of intra- and inter-orbital pairing amplitudes may appear in the superconducting state of the considered model. However, the dominant one corresponds to the nearest neighbor $d$-$d$ pairing. The pairing amplitudes which are going to be analyzed here are denoted by $\Delta_{ll'} \equiv \langle\Psi_G|\hat{c}^{\dagger}_{il\uparrow}\hat{c}^{\dagger}_{jl'\downarrow}|\Psi_G\rangle/\langle\Psi_G|\Psi_G\rangle$. In this analysis we have taken into account the nearest neighbor $d$-$d$ pairing ($\Delta_{dd}$), the nearest neighbor $d$-$p$ pairing ($\Delta_{dp}$), as well as the next nearest-neighbor $p$-$p$ pairing denoted by $\Delta^{||}_{pp}$ and $\Delta^{\perp}_{pp}$ depending on whether the atomic sites between which the pairing occurs lay along the direction distinguished by the $p$ orbital itself or perpendicular to it (cf. \cite{Zegrodnik_3band}).

As shown in Refs. \cite{Kaczmarczyk_2013,Kaczmarczyk2014}, the minimization condition of the system energy in the state $|\Psi_G\rangle$ is equivalent to solving the Sch\"odinger equation with an effective Hamiltonian, which detailed form is provided in Ref. \cite{Zegrodnik_3band} for the case of the three band $d$-$p$ model. By using the dispersion relations in the upper hybridized band of the effective Hamiltonian, $\epsilon^{\mathrm{eff}}(\mathbf{k})$, one can extract the nodal characteristics of the system such as the Fermi momentum $\mathbf{k}_F$, Fermi velocity $v_F=\nabla_{\mathbf{k}}\epsilon^{\mathrm{eff}}(\mathbf{k})|_{\mathbf{k}=\mathbf{k}_F}$, and effective mass, $m^{\mathrm{eff}}$, which are to be analyzed below.

\section{Results and discussion}
In all the presented results we adopt the convention, where zero doping ($\delta=0$) case corresponds to the parent compound for which each CuO$_2$ complex is occupied by five electrons ($n_{tot}=5$). Such situation is referred to as the half-filling, while the hole doped situation $\delta>0$ refers to $n_{tot}<5$.

\subsection{Nodal characteristics in the paired state and comparison to experiment}
Here, we utilize typical hopping and interaction parameters which correspond to the cuprates: $t_{dp}=1.0\;$eV, $t_{pp}=0.4\;$eV, $\epsilon_{dp}=3.2\;$eV, $U_d=11\;$eV, $U_p=4.1\;$eV and analyze first the nodal direction characteristics of the system. At this stage, the next nearest-neighbor, $p$-$p$ electron hopping, and the intersite Coulomb repulsion terms are omitted and are taken into account in the next subsection.

\begin{figure}
\centering
\includegraphics[width=0.6\textwidth]{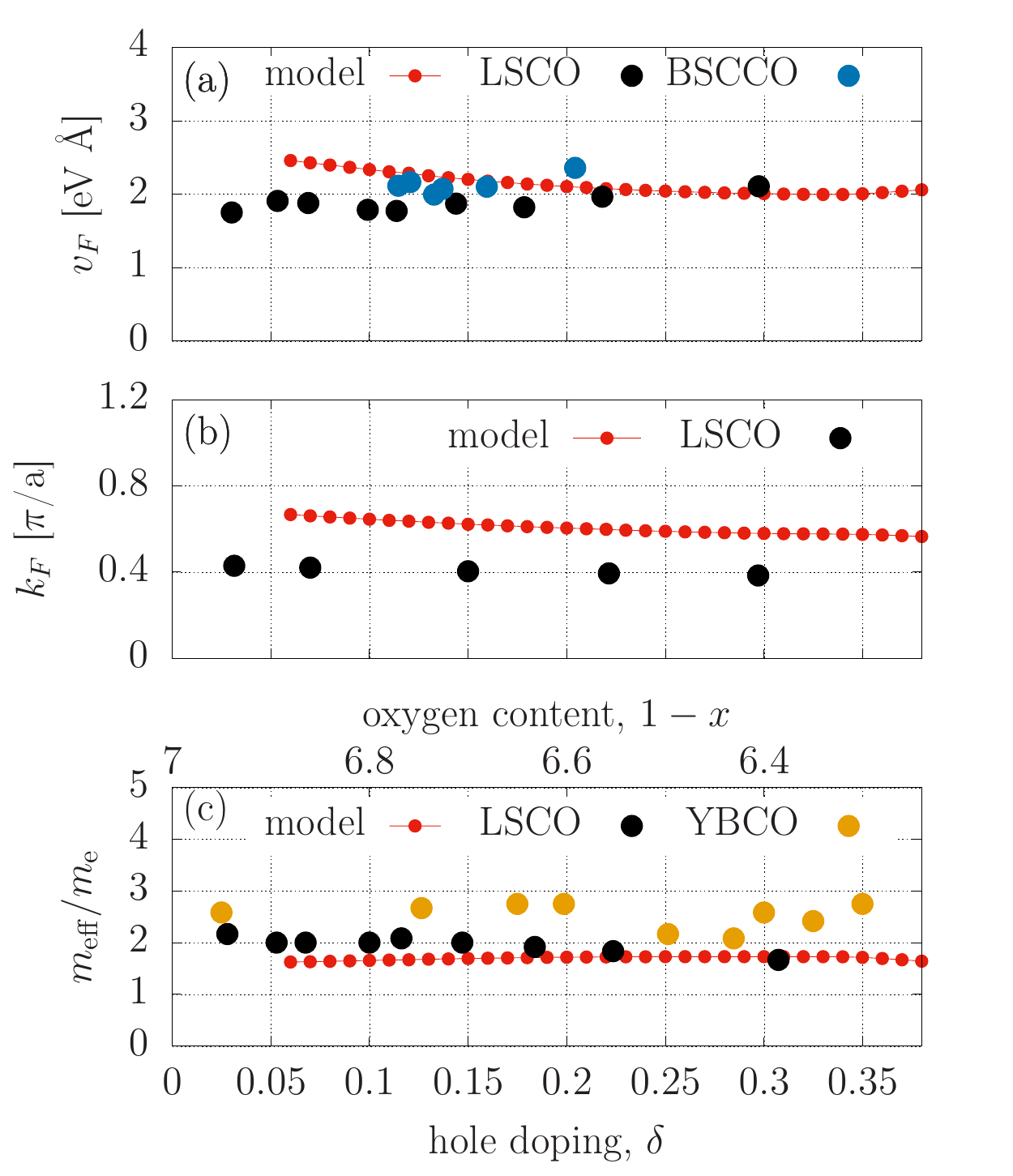}
\caption{(a) Nodal Fermi velocity as a function of hole doping obtained in the three-band $d$-$p$ model with the use of the DE-GWF method, compared with the available experimental data for the LSCO and BSCCO compounds taken from Refs. \cite{Zhou2003} and \cite{Kordyuk2005}, respectively; (b) Theoretical Fermi momentum as a function of hole doping with the corresponding experimental values for LSCO taken from Ref. \cite{Hashimoto2008}; (c) The theoretical effective mass in the units of electron mass as a function of hole doping compared with the corresponding experimental values. The effective mass for LSCO was calculated by using the measured Fermi velocity and Fermi momentum taken from Refs. \cite{Zhou2003} and \cite{Hashimoto2008}, respectively. The experimental values for YBCO are taken from Ref. \cite{Padilla2005} with the doping values determined by the oxygen content (top axis) from the original data.}
\label{fig:nodal}
\end{figure}

In Fig. \ref{fig:nodal} we show the calculated Fermi velocity, Fermi momentum, and effective mass, all as a function of hole doping. For comparison the available experimental data for La$_{2-x}$Sr$_x$CuO (LSCO) and Bi$_2$Sr$_2$CaCu$_2$O$_8$ (BSCCO), and YBa$_2$Cu$_3$O$_7$ (YBCO), taken from 
Refs. \cite{Zhou2003,Kordyuk2005,Padilla2005}, are also shown in the Figure. When it comes to quantitative comparison, a systematic difference between experiment and theory is observed for the case of Fermi momentum. However, the very weak doping dependence of all the calculated nodal characteristics is reproduced well and for a wide doping range the quantitative agreement has been obtained for the case of both Fermi velocity and effective mass. The analysis of the nodal direction features of the cuprates carried out with the use of single-band $t$-$J$-$U$ model \cite{Zegrodnik_1}, leads to similar results, though the quantitative agreement with experimental values obtained there is of better quality. 

\begin{figure}
\centering
\includegraphics[width=0.6\textwidth]{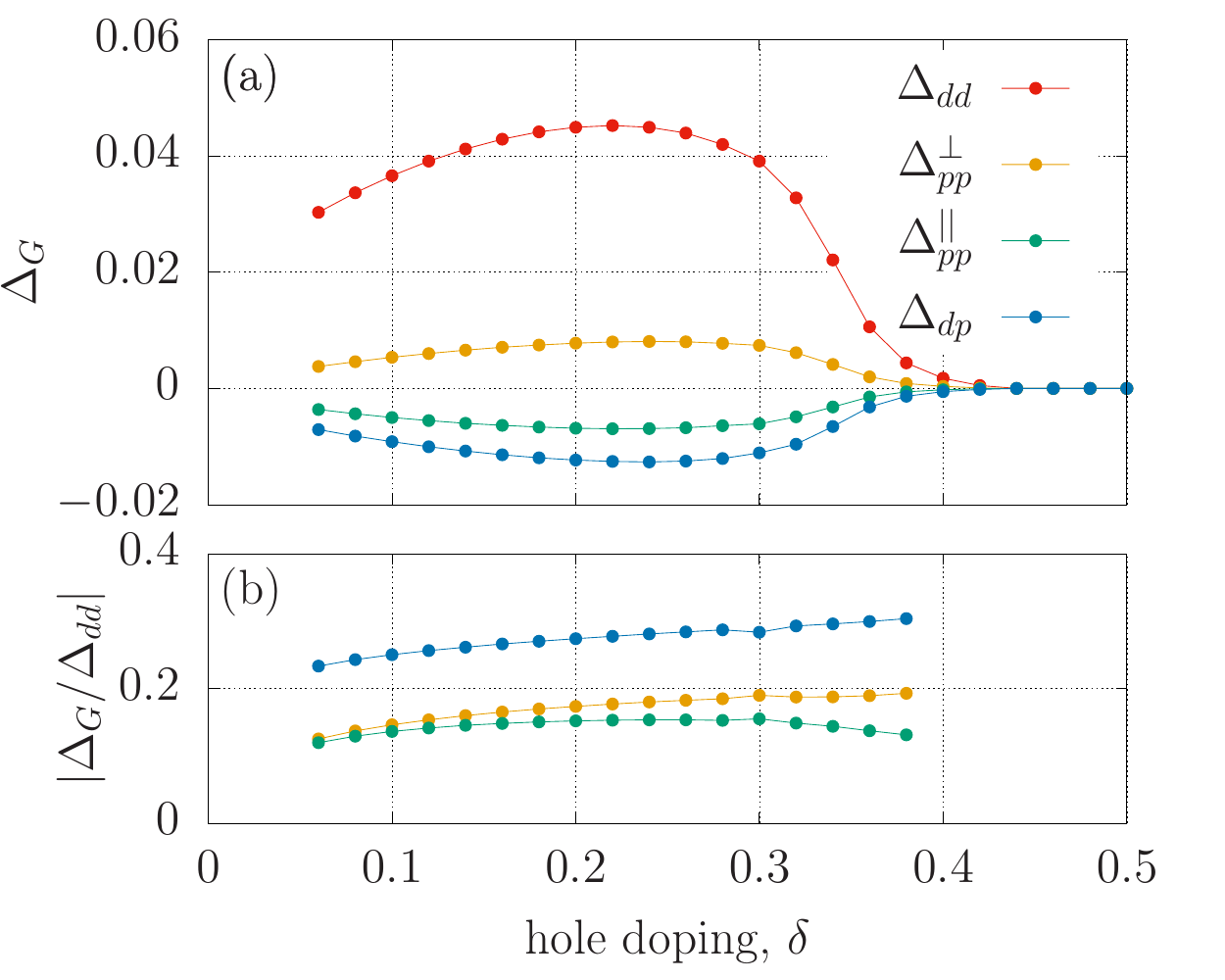}
\caption{(a) The correlated $d$-$d$, $p$-$p$ and $d$-$p$ supercondcting gaps as a function of hole doping ($t_{pp}=V_{pp}=V_{dp}=V_{dd}=0$); (b) The $\Delta^{||}_{pp}$, $\Delta^{\perp}_{pp}$, and $\Delta_{dp}$ gap amplitudes divided by the dominant $d$-$d$ contribution to the pairing.}
\label{fig:delta}
\end{figure}

For the sake of completeness, the calculated doping dependences of the pairing amplitudes are provided in Fig. \ref{fig:delta}. As one can see, the dominant pairing amplitude corresponds to the nearest neighbor $d$-$d$ pairing. As shown in Fig. \ref{fig:delta} (b), the quotient between the remaining pairing amplitudes and $\Delta_{dd}$ is relatively small and weakly doping dependent. Therefore, $T_C$, as well as the shape of the SC phase stability regime on the cuprate phase diagram, is determined mainly by $\Delta_{dd}$. However, the values of the optimal doping and the upper critical doping determined with the use of the $\Delta_{dd}$ plot from Fig. \ref{fig:delta} (a) are overestimated by $\approx 25\%$ with respect to the experimental data \cite{Hufner2008}. As we show in the following subsections, both $t_{pp}$ and $V_{dd}$ reduce the theoretically determined values leading to better agreement between theory and experiment. In spite of significant changes in the values of the pairing amplitudes for the doping range $0-0.35$, the nodal direction features remain well established and weakly doping dependent (cf. Fig. \ref{fig:nodal}), which is partly due to the $d$-$wave$ symmetry of the superconducting gap. It should be noted that, apart from the fact that the paired state is of the  $d$-$wave$ symmetry, an important ingredient leading to the reproduced behavior of the nodal features is the electron-electron repulsion taken into account on sufficiently high accuracy level within the DE-GWF approach.



\subsection{Influence of the next-nearest neighbor $p$-$p$ hopping and intersite Coulomb repulsion}

Here, we consider the case of non-zero next-nearest neighbor oxygen-oxygen hopping term, as well as the intersite $d$-$p$, $p_x$-$p_y$, and $d$-$d$ Coulomb repulsion terms. The selected values of the model parameters, corresponding to the additional terms in Hamiltonian, are provided in each particular situation considered. The values of the remaining parameters are: $t_{dp}=1.14\;$eV, $t_{pp}=0.53\;$eV, $\epsilon_{dp}=3.39\;$eV, $U_d=10.5\;$eV, $U_p=4.87\;$eV.

\begin{figure}
\centering
\includegraphics[width=0.6\textwidth]{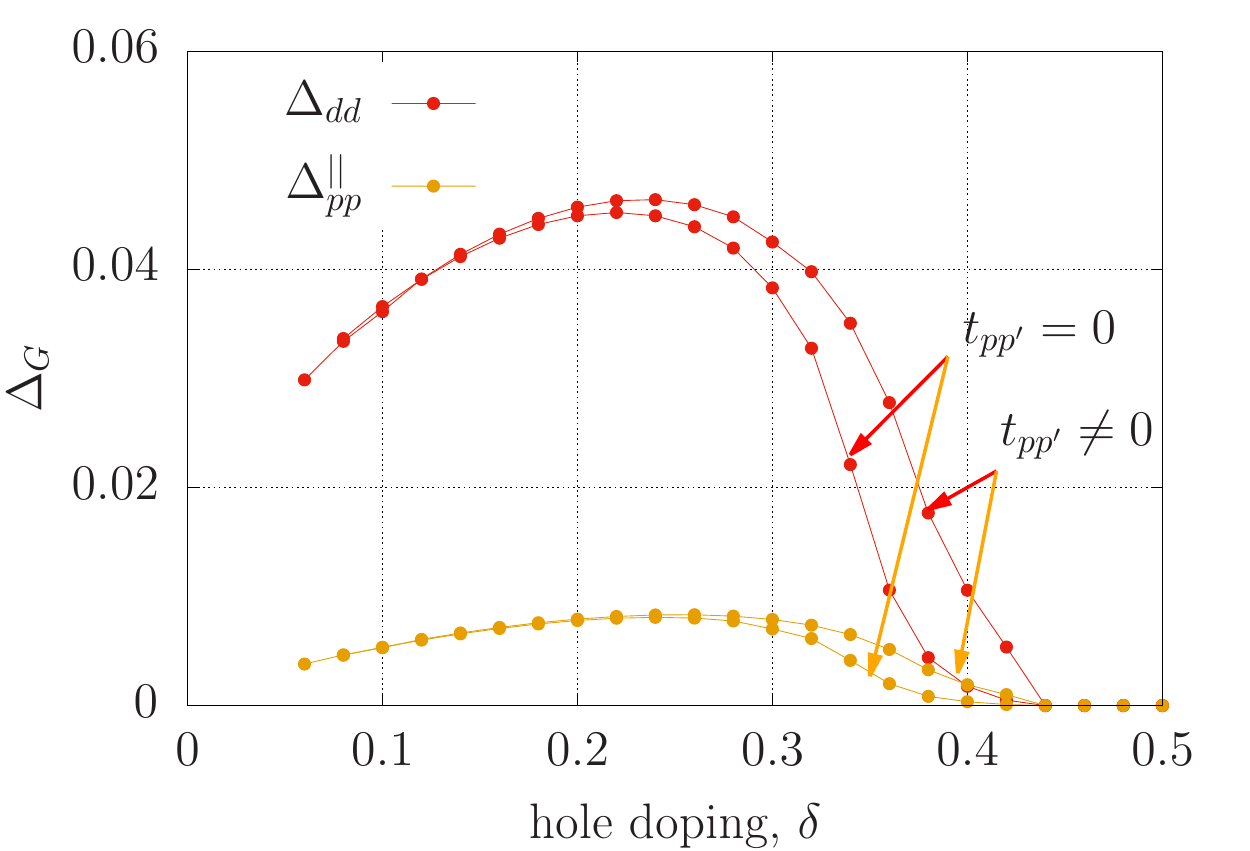}
\caption{(a) The correlated $d$-$d$ and $p$-$p$ supercondcting gaps as a function of hole doping for the case of $t_{pp^{\prime}}=0$ and $t_{pp^{\prime}}=0.2\;$eV. For the sake of clarity we do not show $\Delta^{\perp}_{pp}$ and $\Delta_{dp}$. The effect of $t_{pp}$ on the two latter pairing amplitudes is analogical to the one  shown in the Figure for the case of $\Delta_{dd}$ and $\Delta^{||}_{pp}$.}
\label{fig:delta_tpp}
\end{figure}

\begin{figure}
\centering
\includegraphics[width=0.6\textwidth]{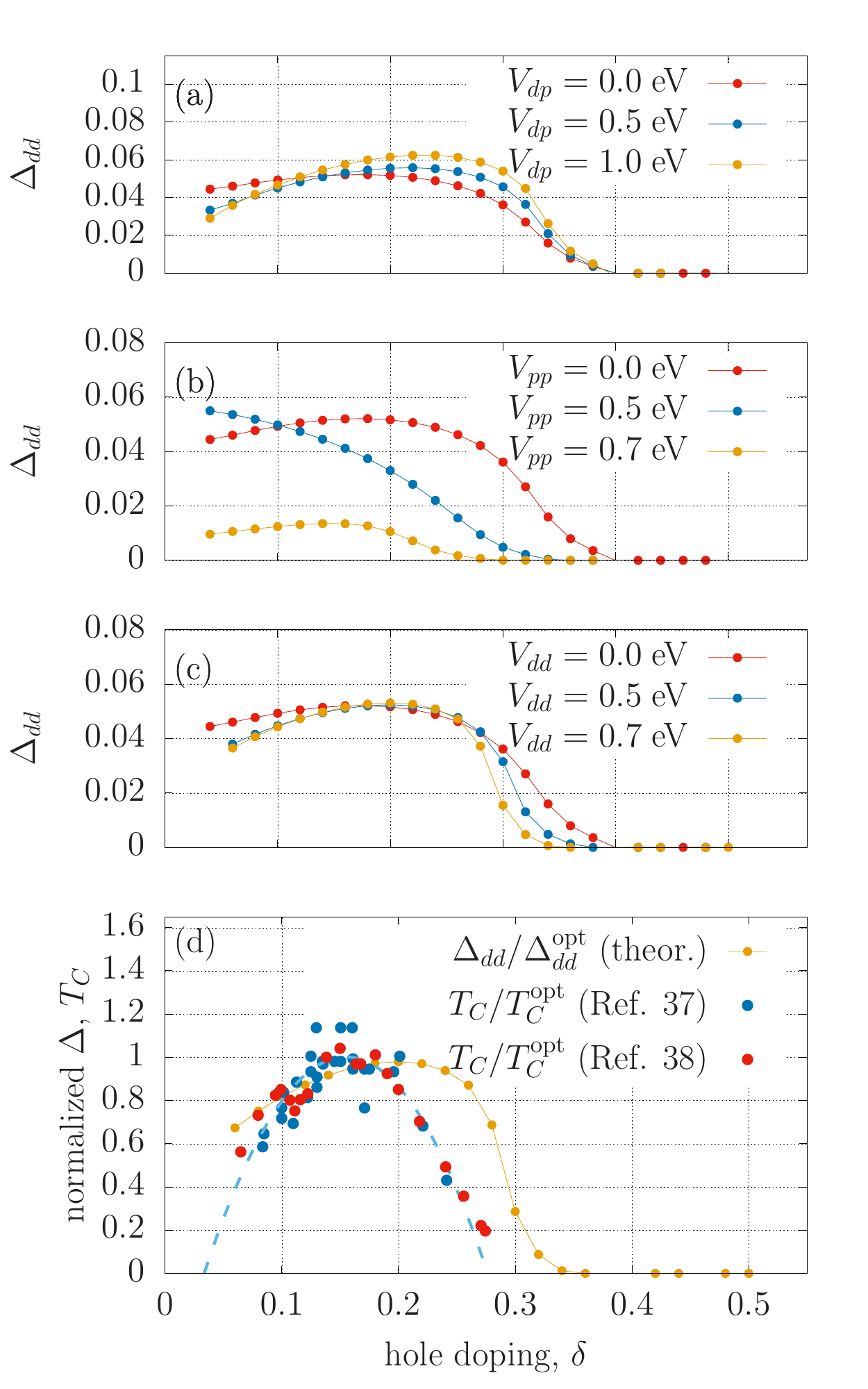}
\caption{The correlated $d$-$d$ supercondcting gaps as a function of hole doping with the inclusion of particular inter-site Coulomb repulsion terms (for $t_{pp}=0$). In (a), (b), and (c) we show the results for nonzero $V_{dp}$, $V_{pp}$, and $V_{dd}$ parameters, respectively. In (d) we compare the theoretical result for $V_{dd}=0.7\;$eV with the experimental data for various systems from the cuprate family. Blue dots represent data for Bi2212, Y123, Tl2201, and Hg120 compounds taken from Ref. \cite{Hufner2008} and red dots represent data for LSCO taken from Ref. \cite{Mollo2003}. The dashed blue line in (d) is a guide to the eye showing the trend of the experimental results.}
\label{fig:delta_V}
\end{figure}

In general, the value of $t_{pp^{\prime}}$ for the compounds from the cuprate family ranges between $\sim 0.1-0.2\;$eV and is believed to be correlated with the maximal critical temperature \cite{Weber_epl2012}. Namely, for larger values of $t_{pp^{\prime}}$ higher $T^{\mathrm{max}}_{C}$ should appear. In Fig. \ref{fig:delta_tpp} we show the calculated $d$-$d$ and $p$-$p$ pairing amplitudes for the case of $t_{pp^{\prime}}=0$ and $t_{pp^{\prime}}=0.2\;$eV. The behavior of the pairing amplitudes $\Delta_{dp}$ and $\Delta^{||}_{pp}$ is very similar to the one shown in Fig. \ref{fig:delta_tpp}. However, we do not show them here for the sake of clarity. As one can see, within our theoretical approach the inclusion of the next-nearest-neighbor $p$-$p$ hopping enhances only slightly the superconducting state by increasing the maximal value of the gap parameters and widening the SC stability doping range. According to the experimental analysis, for the range of $t_{pp^{\prime}}$ values considered here, significant changes of $T^{\mathrm{max}}_{C}$ should be observed. If the correlated pairing amplitude is proportional to $T_C$, the latter observation would be in contradiction to the result presented in Fig. \ref{fig:delta_tpp}. However, as discussed in Ref. \cite{Weber_epl2012}, the value of $t_{pp^{\prime}}$ is directly related to the distance between the apical oxygen atoms and the Cu-O plane. Therefore, the observed changes of $T^{\mathrm{max}}_{C}$ may in fact not be caused by the $t_{pp^{\prime}}$ itself, but instead by more complex effects related to the apical oxygen appearance in close proximity of the copper-oxide plane, what may lead to, e.g., modification of the in-plane exchange interaction \cite{Peng2017}.

\begin{figure}[h!]
\centering
\includegraphics[width=0.6\textwidth]{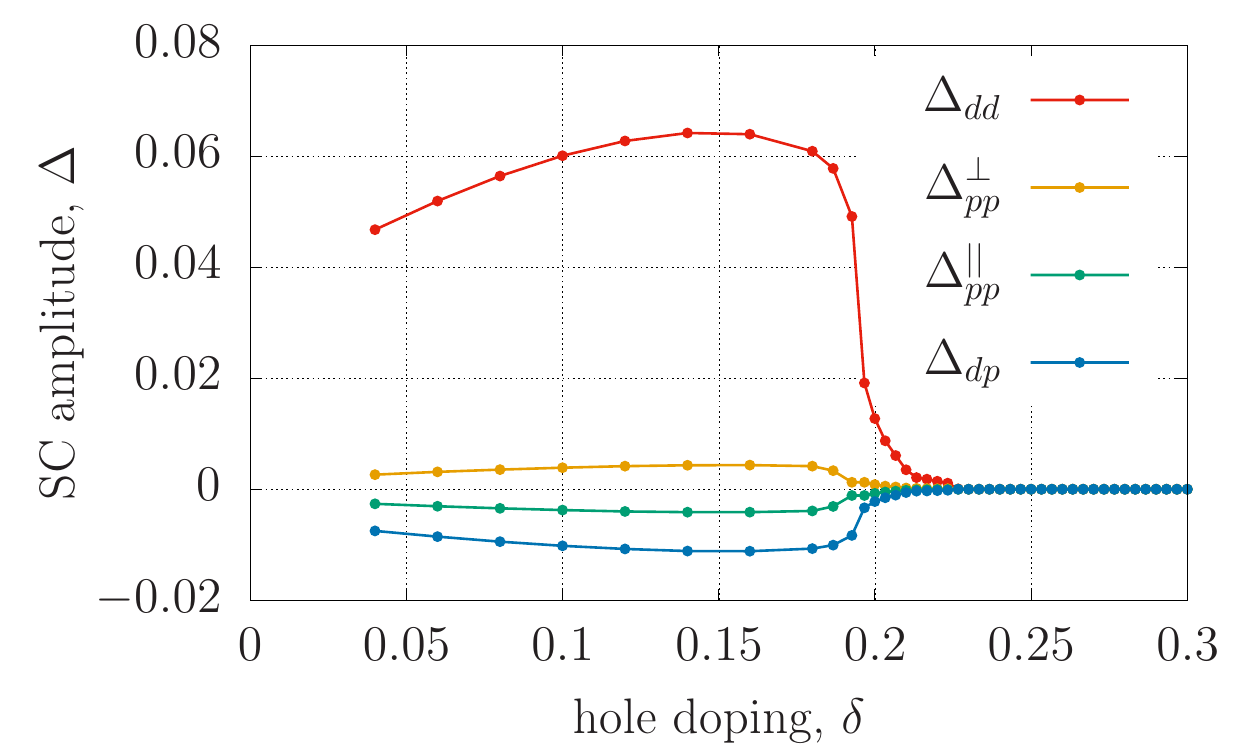}
\caption{The correlated $d$-$d$, $p$-$p$ and $d$-$p$ supercondcting gaps as a function of hole doping with the inclusion of all the considered inter-site Coulmb interaction terms. The model parameters correspond to the La$_2$CuO$_4$ compound and have been taken from \cite{Hirayama_GWDFT_2018}.
}
\label{fig:delta_La_V}
\end{figure}

In Fig. \ref{fig:delta_V} we show the influence of the inter-site Coulomb repulsion terms on the $d$-$d$ pairing amplitude, which constitutes the dominant contribution to superconducting state. As can be seen, the $V_{dp}$ term leads to a slight enhancement of the pairing strength. On the other hand, $V_{pp}$ and $V_{dd}$ have negative effect on the latter, decreasing the pairing amplitude and the doping range of the SC stability. The reduction of both the optimal doping and the upper critical doping resulting from the $V_{dd}$ term leads to a better agreement between the experimental superconducting dome and the theoretical results, as shown in Fig. \ref{fig:delta_V} (d). Nevertheless, there is still room for improvement when it comes to the quantitative analysis. Among all the inter-site Coulomb repulsion terms considered here, the most significant effect comes from the nearest-neighbor $p$-$p$ repulsion. Relatively small value of $V_{pp}=0.7\;$eV already suppresses $\Delta_{dd}$ about five times. Also, the superconducting dome-like structure is distorted upon further increasing $V_{pp}$. This effect should be associated with a simple Hartree-Fock contribution corresponding to the $V_{pp}$-term, which apart from the more complex correlation effects is also taken into account within the DE-GWF scheme. Namely, according to the Hartree-Fock approach, electrons occupying a single $p_x$ ($p_y$) orbital experience an increase of energy of $2V_{pp}n_{p_y}$ ($2V_{pp}n_{p_x}$) due to the oxygen-oxygen repulsion. This lifts the effective atomic levels at the oxygen orbitals, what in turn, is equivalent with decreasing the $\epsilon_{dp}$ value in the initial Hamiltonian. As we have shown in Ref. \cite{Zegrodnik_3band} (Fig. 12 of that paper) by reducing significantly $\epsilon_{dp}$ one can suppress the strength of the pairing in the considered model for given $U_d$. In the considered hole doping range $n_{p_x}, n_{p_y}\approx 1.5-2.0$, thus the effective atomic level change at the oxygen orbitals is $\sim 2.1-2.8\;$eV which is already a relatively large value considering that $\epsilon_{dp}\gtrsim3\;$eV.

It should be noted that all the calculated values presented here do not reach precisely the $\delta=0$ limit. This is caused by the fact that we were not able to reach numerical convergence for our multiband model in the region $\delta\approx 0$. Therefore, it is not clear at this point what is the precise behavior of the gap amplitudes close to the zero-doping limit and if they actually drop to zero there. For particular cases shown in Fig. \ref{fig:delta_V}, the extrapolation of the gap to the $\delta=0$ point would show a nonzero value, which is not what one would expect. Most probably, a further increase of the $U_{dd}$ value would suppress completely the pairing amplitudes for the parent compound case ($\delta=0$). The issue of nonzero values of the pairing strength in the zero doping limit even for significant onsite Coulomb repulsion integrals has been discussed for the case of the single-band Hubbard models already some time ago \cite{Gull2012,Kaczmarczyk_2013}, as well as for the case of the three-band $d$-$p$ model in Ref. \cite{Zegrodnik_3band}. 

In connection with the issue of the paired state behavior in the low-doping regime, it should be noted that the antiferromagnetic (AF) phase is expected to appear there together, with a possible narrow region of antiferromagnetic-superconducting coexistence. Such SC+AF state has been reported in the case of  single band models with a competing character of the AF and SC interplay \cite{Pathak2009,Kaczmarczyk2011,Abram2017}. A similar effect is expected here since the dominant contribution to the superconducting state comes from the pairing between the nearest-neighbor copper atomic sites, on which also the staggered magnetic moments reside in the AF phase. Therefore, a simple extrapolation of the correlated gap amplitudes presented here to the zero doping regime would not lead to the true behavior, since the AF state is not taken into account here. The available analysis of the AF and SC phase within the three-band picture show a proper sequence of phases at the phase diagram \cite{Yanagisawa2008,3band_variational_SDW_SC_2,cui2020groundstate,Yanagisawa2020}. However, the upper critical doping for the disappearance of the AF phase seems to be significantly larger than the experimental one.

\begin{figure}
\centering
\includegraphics[width=0.6\textwidth]{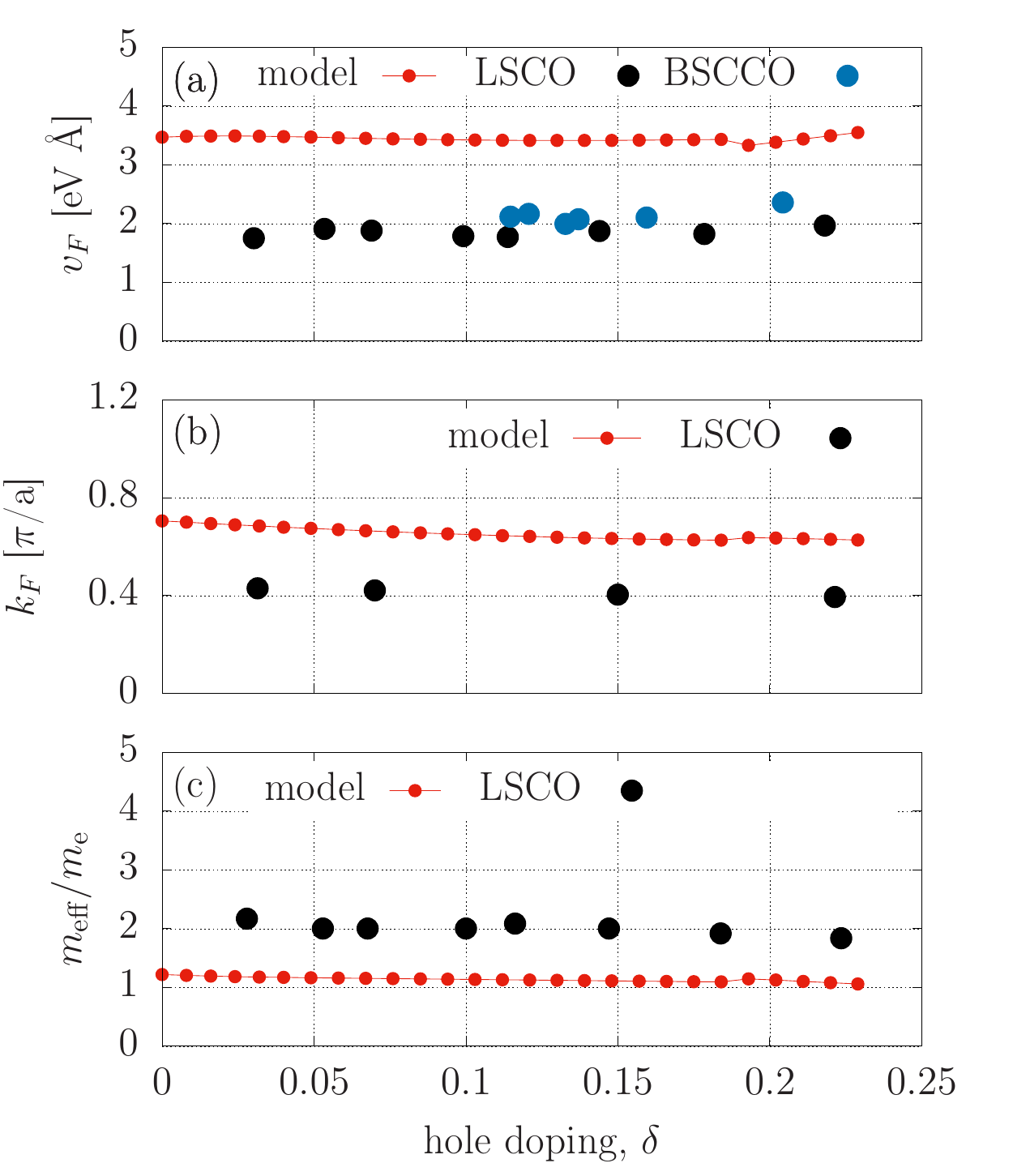}
\caption{Fermi velocity (a), Fermi momentum (b), and effective mass (c) obtained within the extended $d$-$p$ model, with the hopping and interaction parameters corresponding to LSCO and taken from Ref. \cite{Hirayama_GWDFT_2018}. The experimental data provided in the Figure for the Fermi velocity and Fermi momentum (black circles) are taken from Refs. \cite{Zhou2003} and \cite{Hashimoto2008}, respectively. The experimental effective mass for LSCO (black circles) was calculated by using the measured Fermi velocity and Fermi momentum taken from Refs. \cite{Zhou2003} and \cite{Hashimoto2008}, respectively.}
\label{fig:vf_LSCO}
\end{figure}

We have also carried out calculations with the inclusion of all the mentioned terms simultaneously to check if the proper reconstruction of the SC dome is possible in such a general situation. The model parameters have been taken from Ref. \cite{Hirayama_GWDFT_2018}, where a combined $ab$ $initio$ GW and DFT approach has been used, that does not suffer from the so-called double counting problem. The parameters correspond to the La$_2$CuO$_4$ compound and their values are: $t_{dp}=1.369\;$eV, $t_{pp}=0.754\;$eV, $\epsilon_{dp}=3.699\;$eV, $U_d=9.61\;$eV, $U_p=6.13\;$eV, $V_{dd}=1.51\;$eV, $V_{dp}=2.68\;$eV, $V_{pp}=1.86\;$eV. In Fig. \ref{fig:delta_La_V} we show the pairing amplitudes obtained within our DE-GWF calculation scheme. As one can see the SC dome still appears with the maximal critical doping $\delta\gtrsim 0.2$ and the optimal doping $\delta\approx0.15$. Also, the $d$-$d$ pairing amplitude remains the dominant one and relations between particular amplitudes are approximately the same as previously (cf. Fig. \ref{fig:delta}). It should be noted that the obtained here upper critical doping is about 10$\%$ to small in comparison with the experimental phase diagram for LSCO provided in Fig. \ref{fig:delta_V} (d). Therefore, the agreement should be considered as qualitative in this respect. The nodal characteristics calculated for the extended version of the $d$-$p$ model are shown in Fig. \ref{fig:vf_LSCO} and are weakly doping dependent which again qualitatively agrees with the experimental picture. However, the obtained values for $v_F$ and $k_F$ ($m_{\mathrm{eff}}$) are approximately two times larger (smaller) than those measured in the LSCO samples. It is not clear at this stage if the DE-GWF method leads to this discrepancy or the model parameters themselves do not describe properly the LSCO compound \cite{Hirayama_GWDFT_2018}.

\section{Conclusions and outlook}
The calculated nodal direction characteristics obtained within the three-band model approach show a relatively good agreement with the available experimental data as well as with the corresponding single-band calculations. Extended analysis with the inclusion of the inter-site interaction terms show that the $V_{dp}$ term enhances slightly the strength of the pairing, whereas $V_{pp}$ and $V_{dd}$ have negative effect on the paired state, decreasing the pairing amplitude and the doping range of the SC state stability. Among all the inter-site Coulomb repulsion terms considered  here, the most significant effect comes from the nearest-neighbor $p$-$p$ repulsion. Nevertheless, the calculations in the general case with all the mentioned interaction terms taken into account, and the parameters appropriate for LSCO, has lead to the reconstruction of the superconducting dome on the phase diagram and weak doping dependence of the nodal direction characteristics. Also, in all the cases considered here, doping dependencies of the $d$-$d$, $d$-$p$, and $p$-$p$ pairing amplitudes are all very similar. However, the latter two are scaled down by the factor of $\sim 5$ with respect to the first one. Therefore, $T_C$ as well as the shape of the SC phase stability regime on the cuprate phase diagram, will be determined mainly by $\Delta_{dd}$. Our results point to the conclusion that the overall picture of the SC state seems to be reasonably described both in the effective single-band models, as well as different variants of the three-band models. However, the full quantitative theory of high-$T_C$ superconductivity may involve a delicate balance of the possible intra- and interorbital interactions.

\section{Acknowledgement}
MZ and AB acknowledge the financial support through the Grant SONATA, No. 2016/21/D/ST3/00979 from the National Science Centre (NCN), Poland. JS and MF acknowledges  the  financial  support  by the Grant OPUS No. UMO-2018/29/B/ST3/02646 from the National Science Centre (NCN), Poland. This work is supported in part by PL-Grid Infrastructure.

\vspace{20pt}
\bibliographystyle{ieeetr}
\bibliography{refs}

\begin{thebibliography}{10}

\bibitem{cuprates_rev_2006}
P.~A. Lee, N.~Nagaosa, and X.-G. Wen, ``Doping a {M}ott insulator: Physics of
  high-temperature superconductivity,'' {\em Rev. Mod. Phys.}, vol.~78,
  pp.~17--85, Jan 2006.

\bibitem{tJmodel_rev_2008}
M.~Ogata and H.~Fukuyama, ``The $t-{J}$ model for the oxide high ${T}_c$
  superconductors,'' {\em Reports on Progress in Physics}, vol.~71, no.~3,
  p.~036501, 2008.

\bibitem{Anderson1988}
P.~W. Anderson, {\em Frontiers and Borderlines in Many Particle Physics}.
\newblock Amsterdam: North-Holland, 1988.

\bibitem{ZRS1988}
F.~C. Zhang and T.~M. Rice, ``Effective hamiltonian for the superconducting
  {C}u oxides,'' {\em Phys. Rev. B}, vol.~37, pp.~3759--3761, Mar 1988.

\bibitem{Dagotto1994}
E.~Dagotto, ``Correlated electrons in high-temperature superconductors,'' {\em
  Rev. Mod. Phys.}, vol.~66, pp.~763--840, Jul 1994.

\bibitem{VMC_Hubbard_2007}
D.~Eichenberger and D.~Baeriswyl, ``Superconductivity and antiferromagnetism in
  the two-dimensional {H}ubbard model: A variational study,'' {\em Phys. Rev.
  B}, vol.~76, p.~180504, Nov 2007.

\bibitem{DMFT_SC_Hub_2006}
M.~Capone and G.~Kotliar, ``Competition between $d$-wave superconductivity and
  antiferromagnetism in the two-dimensional {H}ubbard model,'' {\em Phys. Rev.
  B}, vol.~74, p.~054513, Aug 2006.

\bibitem{Hub_SC_2005}
D.~S\'en\'echal, P.-L. Lavertu, M.-A. Marois, and A.-M.~S. Tremblay,
  ``Competition between antiferromagnetism and superconductivity in
  high-${T}_{c}$ cuprates,'' {\em Phys. Rev. Lett.}, vol.~94, p.~156404, Apr
  2005.

\bibitem{Gros2007}
B.~Edegger, V.~N. Muthukumar, and C.~Gros, ``Gutzwiller–rvb theory of
  high-temperature superconductivity: Results from renormalized mean-field
  theory and variational monte carlo calculations,'' {\em Advances in Physics},
  vol.~56, no.~6, pp.~927--1033, 2007.

\bibitem{Fidrysiak2020}
M.~Fidrysiak and J.~Spa\l{}ek, ``Robust spin and charge excitations throughout
  the high-${T}_{c}$ cuprate phase diagram from incipient mottness,'' {\em
  Phys. Rev. B}, vol.~102, p.~014505, Jul 2020.

\bibitem{Zegrodnik_1}
J.~Spa\l{}ek, M.~Zegrodnik, and J.~Kaczmarczyk, ``Universal properties of
  high-temperature superconductors from real-space pairing:
  $t\ensuremath{-}j\ensuremath{-}u$ model and its quantitative comparison with
  experiment,'' {\em Phys. Rev. B}, vol.~95, p.~024506, Jan 2017.

\bibitem{Zegrodnik_5}
M.~Zegrodnik and J.~Spa\l{}ek, ``Incorporation of charge- and pair-density-wave
  states into the one-band model of $d$-wave superconductivity,'' {\em Phys.
  Rev. B}, vol.~98, p.~155144, Oct 2018.

\bibitem{Fidrysiak2018}
M.~Fidrysiak, M.~Zegrodnik, and J.~Spałek, ``Realistic estimates of
  superconducting properties for the cuprates: reciprocal-space diagrammatic
  expansion combined with variational approach,'' {\em Journal of Physics:
  Condensed Matter}, vol.~30, no.~47, p.~475602, 2018.

\bibitem{3band_PRL_Mott}
A.~Go and A.~J. Millis, ``Spatial correlations and the insulating phase of the
  high-${T}_{c}$ cuprates: Insights from a configuration-interaction-based
  solver for dynamical mean field theory,'' {\em Phys. Rev. Lett.}, vol.~114,
  p.~016402, Jan 2015.

\bibitem{QMC_3band_2016}
Y.~F. Kung, C.-C. Chen, Y.~Wang, E.~W. Huang, E.~A. Nowadnick, B.~Moritz, R.~T.
  Scalettar, S.~Johnston, and T.~P. Devereaux, ``Characterizing the
  three-orbital {H}ubbard model with determinant quantum monte carlo,'' {\em
  Phys. Rev. B}, vol.~93, p.~155166, Apr 2016.

\bibitem{Yanagisawa2008}
T.~Yanagisawa, M.~Miyazaki, and K.~Yamaji, ``Incommensurate antiferromagnetism
  coexisting with superconductivity in two-dimensional d–p model,'' {\em
  Journal of the Physical Society of Japan}, vol.~78, no.~1, p.~013706, 2009.

\bibitem{Weber_epl2012}
C.~Weber, C.~Yee, K.~Haule, and G.~Kotliar, ``Scaling of the transition
  temperature of hole-doped cuprate superconductors with the charge-transfer
  energy,'' {\em EPL (Europhysics Letters)}, vol.~100, no.~3, p.~37001, 2012.

\bibitem{Zegrodnik_3band}
M.~Zegrodnik, A.~Biborski, M.~Fidrysiak, and J.~Spa\l{}ek, ``Superconductivity
  in the three-band model of cuprates: Variational wave function study and
  relation to the single-band case,'' {\em Phys. Rev. B}, vol.~99, p.~104511,
  Mar 2019.

\bibitem{Zheng1995}
G.-q. Zheng, Y.~Kitaoka, K.~Ishida, and K.~Asayama, ``Local hole distribution
  in the {C}u{O}$_2$ plane of high-${T}_c$ {C}u-oxides studied by {C}u and
  oxygen {NQR}/{NMR},'' {\em Journal of the Physical Society of Japan},
  vol.~64, no.~7, pp.~2524--2532, 1995.

\bibitem{Rybicki2016}
D.~Rybicki, M.~Jurkutat, S.~Reichardt, C.~Kapusta, and J.~Haase, ``Perspective
  on the phase diagram of cuprate high-temperature superconductors,'' {\em Nat.
  Commun.}, vol.~7, p.~11413, 2016.

\bibitem{Ruan_Sci_Biull2016}
W.~Ruan, C.~Hu, J.~Zhao, P.~Cai, Y.~Peng, C.~Ye, R.~Yu, X.~Li, Z.~Hao, C.~Jin,
  X.~Zhou, Z.-Y. Weng, and Y.~Wang, ``Relationship between the parent charge
  transfer gap and maximum transition temperature in cuprates,'' {\em Science
  Bulletin}, vol.~61, no.~23, pp.~1826 -- 1832, 2016.

\bibitem{biborski2020}
A.~Biborski, M.~Zegrodnik, and J.~Spa\l{}ek, ``Superconducting properties of
  the hole-doped three-band $d\ensuremath{-}p$ model studied with minimal-size
  real-space $d$-wave pairing operators,'' {\em Phys. Rev. B}, vol.~101,
  p.~214504, Jun 2020.

\bibitem{Zegrodnik_nematic}
M.~Zegrodnik, A.~Biborski, and J.~Spa\l{}ek, ``Superconductivity and
  intra-unit-cell electronic nematic phase in the three-bandmodel of
  cuprates,'' {\em Eur. Phys. J. B}, vol.~93, p.~183, 2020.

\bibitem{Peng2017}
Y.~Y. Peng, G.~Dellea, M.~Minola, M.~Conni, A.~Amorese, D.~D. Castro, G.~M.~D.
  Luca, K.~Kummer, M.~Salluzzo, X.~Sun, X.~J. Zhou, G.~Balestrino, M.~L. Tacon,
  B.~Keimer, L.~Braicovich, N.~B. Brookes, and G.~Ghiringhelli, ``Influence of
  apical oxygen on the extent of in-plane exchange interaction in cuprate
  superconductors,'' {\em Nature Physics}, vol.~13, p.~1201–1206, 2017.

\bibitem{Pavarini}
E.~Pavarini, I.~Dasgupta, T.~Saha-Dasgupta, O.~Jepsen, and O.~K. Andersen,
  ``Band-structure trend in hole-doped cuprates and correlation with
  ${T}_{\mathit{c}\mathrm{max}}$,'' {\em Phys. Rev. Lett.}, vol.~87, p.~047003,
  Jul 2001.

\bibitem{Hepting}
M.~Hepting, L.~Chaix, E.~W. Huang, R.~Fumagalli, Y.~Y. Peng, B.~Moritz,
  K.~Kummer, N.~B. Brookes, W.~C. Lee, M.~Hashimoto, T.~Sarkar, J.-F. He, C.~R.
  Rotundu, Y.~S. Lee, R.~L. Greene, L.~Braicovich, G.~Ghiringhelli, Z.~X. Shen,
  T.~P. Devereaux, and W.~S. Lee, ``Three-dimensional collective charge
  excitations in electron-doped copper oxide superconductors,'' {\em Nature},
  vol.~536, p.~374, 2018.

\bibitem{Greco}
A.~Greco, H.~Yamase, and M.~Bejas, ``Origin of high-energy charge excitations
  observed by resonant inelastic {X}-ray scattering in cuprate
  superconductors,'' {\em Commun. Phys.}, vol.~2, p.~3, 2019.

\bibitem{Hirayama_GWDFT_2018}
M.~Hirayama, Y.~Yamaji, T.~Misawa, and M.~Imada, ``Ab initio effective
  hamiltonians for cuprate superconductors,'' {\em Phys. Rev. B}, vol.~98,
  p.~134501, Oct 2018.

\bibitem{Kaczmarczyk_2013}
J.~Kaczmarczyk, J.~Spa\l{}ek, T.~Schickling, and J.~B\"unemann,
  ``Superconductivity in the two-dimensional {H}ubbard model: {G}utzwiller wave
  function solution,'' {\em Phys. Rev. B}, vol.~88, p.~115127, Sep 2013.

\bibitem{MWysokinski_Anderson}
M.~M. Wysoki\ifmmode~\acute{n}\else \'{n}\fi{}ski, J.~Kaczmarczyk, and
  J.~Spa\l{}ek, ``Correlation-driven $d$-wave superconductivity in anderson
  lattice model: Two gaps,'' {\em Phys. Rev. B}, vol.~94, p.~024517, Jul 2016.

\bibitem{2band_Hub_Bunemann}
K.~z. M\"unster and J.~B\"unemann, ``Gutzwiller variational wave function for
  multiorbital {H}ubbard models in finite dimensions,'' {\em Phys. Rev. B},
  vol.~94, p.~045135, Jul 2016.

\bibitem{Kaczmarczyk2014}
J.~Kaczmarczyk, J.~Bünemann, and J.~Spałek, ``High-temperature
  superconductivity in the two-dimensional t – {J} model: {G}utzwiller
  wavefunction solution,'' {\em New Journal of Physics}, vol.~16, no.~7,
  p.~073018, 2014.

\bibitem{Zhou2003}
X.~J. Zhou, T.~Yoshida, A.~Lanzara, P.~V. Bogdanov, S.~A. Kellar, K.~M. Shen,
  W.~L. Yang, F.~Ronning, T.~Sasagawa, T.~Kakeshita, T.~Noda, S.~Eisaki,
  H.~Uchida, C.~T. Lin, F.~Zhou, J.~W. Xiong, W.~X. Ti, Z.~X. Zhao,
  A.~Fujimori, Z.~Hussain, and Z.-X. Shen, ``Universal nodal {F}ermi
  velocity,'' {\em Nature}, vol.~423, p.~398, 2003.

\bibitem{Kordyuk2005}
A.~A. Kordyuk, S.~V. Borisenko, A.~Koitzsch, J.~Fink, M.~Knupfer, and
  H.~Berger, ``Bare electron dispersion from experiment: Self-consistent
  self-energy analysis of photoemission data,'' {\em Phys. Rev. B}, vol.~71,
  p.~214513, Jun 2005.

\bibitem{Hashimoto2008}
M.~Hashimoto, T.~Yoshida, H.~Yagi, M.~Takizawa, A.~Fujimori, M.~Kubota, K.~Ono,
  K.~Tanaka, D.~H. Lu, Z.-X. Shen, S.~Ono, and Y.~Ando, ``Doping evolution of
  the electronic structure in the single-layer cuprate
  ${Bi}_{2}{Sr}_{2\ensuremath{-}x}{La}_{x}{C}u{O}_{6+\ensuremath{\delta}}$:
  Comparison with other single-layer cuprates,'' {\em Phys. Rev. B}, vol.~77,
  p.~094516, Mar 2008.

\bibitem{Padilla2005}
W.~J. Padilla, Y.~S. Lee, M.~Dumm, G.~Blumberg, S.~Ono, K.~Segawa, S.~Komiya,
  Y.~Ando, and D.~N. Basov, ``Constant effective mass across the phase diagram
  of high-${T}_{c}$ cuprates,'' {\em Phys. Rev. B}, vol.~72, p.~060511, Aug
  2005.

\bibitem{Hufner2008}
S.~Hüfner, M.~A. Hossain, A.~Damascelli, and G.~A. Sawatzky, ``Two gaps make a
  high-temperature superconductor?,'' {\em Reports on Progress in Physics},
  vol.~71, p.~062501, may 2008.

\bibitem{Mollo2003}
E.~V.~L. de~Mello, E.~S. Caixeiro, and J.~L. Gonz\'alez, ``Theory for
  high-${T}_{c}$ superconductors considering inhomogeneous charge
  distribution,'' {\em Phys. Rev. B}, vol.~67, p.~024502, Jan 2003.

\bibitem{Gull2012}
E.~Gull and A.~J. Millis, ``Energetics of superconductivity in the
  two-dimensional {H}ubbard model,'' {\em Phys. Rev. B}, vol.~86, p.~241106,
  Dec 2012.

\bibitem{Pathak2009}
S.~Pathak, V.~B. Shenoy, M.~Randeria, and N.~Trivedi, ``Competition between
  antiferromagnetic and superconducting states, electron-hole doping asymmetry,
  and {F}ermi-surface topology in high temperature superconductors,'' {\em
  Phys. Rev. Lett.}, vol.~102, p.~027002, Jan 2009.

\bibitem{Kaczmarczyk2011}
J.~Kaczmarczyk and J.~Spa\l{}ek, ``Coexistence of antiferromagnetism and
  superconductivity within $t$-${J}$ model with strong correlations and nonzero
  spin polarization,'' {\em Phys. Rev. B}, vol.~84, p.~125140, Sep 2011.

\bibitem{Abram2017}
M.~Abram, M.~Zegrodnik, and J.~Spałek, ``Antiferromagnetism, charge density
  wave, and d -wave superconductivity in the extended t - {J} - {U} model: role
  of intersite coulomb interaction and a critical overview of renormalized mean
  field theory,'' {\em Journal of Physics: Condensed Matter}, vol.~29, no.~36,
  p.~365602, 2017.

\bibitem{3band_variational_SDW_SC_2}
C.~Weber, T.~Giamarchi, and C.~M. Varma, ``Phase diagram of a three-orbital
  model for high-${T}_{c}$ cuprate superconductors,'' {\em Phys. Rev. Lett.},
  vol.~112, p.~117001, Mar 2014.

\bibitem{cui2020groundstate}
Z.-H. Cui, C.~Sun, U.~Ray, B.-X. Zheng, Q.~Sun, and G.~K.-L. Chan,
  ``Ground-state phase diagram of the three-band {H}ubbard model from density
  matrix embedding theory,'' 2020.

\bibitem{Yanagisawa2020}
T.~Yanagisawa, M.~Miyazaki, and K.~Yamaji, ``Phase diagram of cuprate
  high-temperature superconductors based on the optimization monte carlo
  method,'' {\em Modern Physics Letters B}, vol.~34, p.~2040046, 2020.

\end{thebibliography}


\end{document}